\documentclass[aps,twocolumn,showpacs]{revtex4}
\usepackage{graphicx}
\usepackage{amsmath}

\begin{document}

\title{Semiclassical quantization with bifurcating orbits}
\author{Thomas Bartsch}
\author{J\"org Main}
\author{G\"unter Wunner}
\affiliation{Institut f\"ur Theoretische Physik 1, Universit\"at Stuttgart,
	 70550 Stuttgart, Germany}

\date{\today}

\begin{abstract}
Bifurcations of classical orbits introduce divergences into semiclassical
spectra which have to be smoothed with the help of uniform
approximations. We develop a technique to extract individual energy levels
from semiclassical spectra involving uniform approximations.
As a prototype example, the method is shown to yield excellent
results for photo-absorption spectra for the hydrogen atom in an electric
field in a spectral range where the abundance of bifurcations would render
the standard closed-orbit formula without uniform approximations useless.
Our method immediately applies to semiclassical trace formulae as well as
closed-orbit theory and offers a general technique for the semiclassical
quantization of arbitrary systems.
\end{abstract}


\maketitle

The correspondence between atomic spectra and classical orbits has been of
fundamental interest and importance since the early days of quantum mechanics.
The ``old'' quantum theory suffered from the severe drawbacks that the
Bohr-Sommerfeld quantization rules can only be applied to integrable systems,
and, for atomic systems, the Heisenberg principle for matrix elements is
silent about transition amplitudes between low-lying and highly excited states.
An important step towards a deeper understanding of the intimate connection 
between classical orbits and the quantum spectra was achieved by the
development of periodic orbit theory \cite{Gut90} and, as a variant
for the photo-excitation of atomic systems, closed orbit theory
\cite{Du88,Bog89}.
In these theories, the density of states or photo-absorption spectra are 
given as the sum of two terms, one a smoothly varying part (as a function 
of energy) and the other a superposition of sinusoidal modulations.
The frequencies, amplitudes, and phases of the modulations are directly
given in terms of classical parameters of the orbits.

Closed-orbit theory has proven  a powerful tool for the
semiclassical interpretation of quantum spectra of, e.g., atoms in external 
magnetic and electric fields by explaining the peaks in the Fourier-transform 
recurrence spectra -- qualitatively and even quantitatively -- in terms of 
the closed orbits of the underlying classical system \cite{Mai94,Cou95,Kip99}.
However, more than a decade after the development of closed-orbit theory the
inverse procedure, i.e., the semiclassical calculation of the eigenenergies 
and transition amplitudes of individual states is still an unsolved 
problem.
The reasons are twofold:
Firstly, both closed-orbit and periodic-orbit theory suffer from fundamental 
convergence problems of the infinite orbit sums.
Secondly, in generic systems the orbits undergo bifurcations when the
energy is varied, and the semiclassical theories for isolated orbits 
exhibit unphysical divergences at the bifurcation points.
Both problems have been addressed separately:
Firstly, the harmonic inversion technique was introduced as
a method for semiclassical quantization \cite{Mai97c,Mai99d}, which allows 
one to overcome the convergence problems of the closed orbit sum and to 
extract high-resolution spectra from a finite set of classical orbits.
Secondly, in the vicinity of bifurcations the semiclassical approximation 
for isolated orbits was replaced with a uniform approximation 
describing all orbits involved in a bifurcation collectively
\cite{Mai97a,Gao97}.

In this Letter both ideas are combined for the first time, i.e., we use both
uniform approximations and harmonic inversion techniques for the
semiclassical calculation of high-resolution spectra.  In the presence of
uniform approximations, the classical scaling laws which have been
essential to all previous applications of the harmonic inversion technique
\cite{Mai99d} are no longer valid.  Therefore, the harmonic inversion
method must be
generalized to handle the non-scaling functional form of uniform
semiclassical approximations. It then gains a degree of flexibility in the
quantization of arbitrary systems no other semiclassical quantization
scheme has been able to reach to date.

The novel method will be demonstrated by way of example of the hydrogen
atom in an electric field.  As is well known, the classical dynamics of
this system is integrable, which means that semiclassical energy
eigenvalues can be calculated with the help of the EBKM torus quantization
rules \cite{Gao94,Kon97}. %\cite{Gao94,Kon97,Kon98}
However, when closed-orbit theory is
applied the hydrogen atom in an electric field exhibits properties
typical of mixed regular-chaotic systems as, e.g., Rydberg atoms in a magnetic
field or H\'enon-Heiles type systems.
In particular, the closed orbits starting at and returning to the
nucleus undergo bifurcations as the energy is varied.  Contrary to the
torus quantization, the method introduced in this Letter is not restricted
to the Stark effect but can be applied to a large variety of systems with
chaotic or mixed regular-chaotic classical dynamics. Furthermore, it can be
used in connection with periodic-orbit theory \cite{Gut90} as well as
closed-orbit theory.

The classical dynamics of the Stark system has already been discussed in
detail \cite{Gao94}.  For any energy, the electron can go ``uphill''
against the direction of the electric field until the external field and
the Coulomb field make it return to the nucleus.  Alternatively, the
electron can leave the nucleus in the ``downhill'' direction of the
external field.  The downhill orbit is closed only for energies below the
Stark saddle point energy, $E_{\rm S}=-2F^{1/2}$, otherwise the electron
will cross the Stark saddle and escape to infinity.
In addition to these axial closed orbits, there are non-axial orbits
returning to the nucleus after $k$ oscillations in the downhill direction
and $l>k$ oscillations in the uphill direction.  Each of these orbits is
generated in a bifurcation off the downhill orbit at a critical energy
$E_{\rm gen}$ and destroyed in a collision with the uphill orbit at $E_{\rm
dest} > E_{\rm gen}$.  Outside this energy range, they exist as complex
``ghost'' orbits.

Closed-orbit theory associates modulations observed in the quantum
photo-absorption spectra of Rydberg atoms in external fields with the 
classical closed orbits.
The quantum response function
\begin{equation}
\label{QuantG}
  g(E) = \sum_n \frac{|\langle i|D|n\rangle|^2}{E-E_n+i\epsilon}
       = \langle i|DG_E^+D|i \rangle \;,
\end{equation}
where $|i\rangle$ is the initial state, $D$ the dipole operator and $G_E^+$ 
the retarded Green's function, is given as a smooth background plus an 
oscillatory closed-orbit sum \cite{Du88,Bog89,Gao92}
\begin{equation}
\label{COSum}
  g^{\rm osc}(E) = \sum_{\rm co} {\cal A}_{\rm co}(E) e^{iS_{\rm co}(E)}  \;,
\end{equation}
where $S_{\rm co}$ is the action of a closed orbit (co) and ${\cal A}_{\rm
co}$ a recurrence amplitude calculated from the monodromy matrix of the
orbit and its initial and final directions with respect to the electric
field. It includes a complex phase given by the Maslov index.  In the
following we are using atomic units, with $\hbar=1$ and $F_0=5.14\times
10^9$~V/cm the unit of the electric field strength.

The most convincing semiclassical interpretation of quantum spectra can be
obtained by means of ``scaled energy spectroscopy'':  By rescaling the
classical quantities with suitable powers of, e.g., the electric field
strength $F$, the classical dynamics can be shown not to depend on the
energy $E$ and the field strength $F$ separately, but only on the scaled
energy $\tilde E=E F^{-1/2}$. When recording quantum states at a fixed
scaled energy $\tilde E$ as a function of the scaling parameter
$w=F^{-1/4}$, each isolated closed orbit contributes a sinusoidal modulation 
to the sum (\ref{COSum}), which can be extracted by a Fourier transform of 
the quantum spectrum.
Experimental scaled energy spectra of atoms in electric fields have been
analyzed in this way \cite{Cou95,Kip99}.
The analysis reveals strong evidence for closed orbit bifurcations.

The simple semiclassical approximation embodied in the closed-orbit formula
(\ref{COSum}) fails close to a bifurcation of closed orbits, resulting in
the divergence of the recurrence amplitudes.
To overcome this difficulty, the closed-orbit terms for isolated orbits
in (\ref{COSum}) must be replaced with a uniform approximation describing 
all orbits involved in a bifurcation collectively.
A uniform approximation suitable for regularizing the bifurcation of a 
non-axial orbit off either the downhill or the uphill orbit was derived 
by Gao and Delos \cite{Gao97} as well as Shaw and Robicheaux \cite{Shaw98a}.
We will use a slightly modified version of their result that gives the 
collective contribution of the axial and non-axial orbits participating in 
a bifurcation in terms of their actions $S_{\rm ax}$ and $S_{\rm non}$ and
recurrence amplitudes ${\cal A}_{\rm ax}$ and ${\cal A}_{\rm non}$ as
\begin{equation}
\label{uniform:eq}
  \Psi(E) = 
    \biggl[\frac{{\cal A}_{\rm non}}{(1+i)}\,I 
         +\frac{1}{a}\left(a {\cal A}_{\rm ax}
                           +\frac{1+i}{\sqrt{2\pi}}\,{\cal A}_{\rm non}
                     \right)
    \biggr] e^{i S_{\rm ax}}
\end{equation}
where $I$ is given in terms of the standard Fresnel integrals $C(x)$ and $S(x)$
\cite{Abramowitz},
\begin{equation}
  I=e^{-i a^2/4} 
      \left[\frac{1+i}{2}-C\left(-\frac{a}{\sqrt{2\pi}}\right)
                         -i S\left(-\frac{a}{\sqrt{2\pi}}\right)
      \right] \;,
\end{equation}
and
\begin{equation}
  a=\pm 2\sqrt{S_{\rm ax}-S_{\rm non}} \;.
\end{equation}
The negative sign for $a$ has to be chosen if the non-axial orbit is a 
complex ghost orbit.

The high-resolution quantization by harmonic inversion \cite{Mai97c,Mai99d}
is based on the observation that by equating the quantum recurrence function 
(\ref{QuantG}) to its semiclassical approximation (\ref{COSum}) -- the
smooth part can be neglected -- and taking the Fourier transform we obtain
\begin{equation}
\label{HiAnsatz}
  -i\sum_n d_n e^{-iE_n t} = C(t)
\end{equation}
with $d_n=|\langle i|D|n\rangle|^2$ and 
\begin{equation}
\label{C1}
  C(t) = \frac{1}{2\pi} \int_{-\infty}^{\infty} dE
    \sum_{\rm co} {\cal A}_{\rm co}(E) e^{i S_{\rm co}(E)} e^{-iEt} \;.
\end{equation}
The quantization problem has thus been recast as the problem of extracting
the frequencies $E_n$ and amplitudes $d_n$ from a given time signal $C(t)$
of the form (\ref{HiAnsatz}), provided the signal (\ref{C1}) can be
calculated.  In the case of a scaling system the signal is given as a sum
of $\delta$ functions.

While the uniform approximation (\ref{uniform:eq}) successfully smoothes
the divergences in (\ref{COSum}), it spoils the classical scaling
properties \cite{Shaw98a}.  Therefore, the Fourier transform of a spectrum
including uniform approximations cannot be evaluated in terms of $\delta$
functions. In fact, for non-scaling systems, there seems to be no way at
all to compute the integral (\ref{C1}) because, apart from the fact that
the classical quantities can always be calculated in a finite energy
interval only, the integral can in general not even be expected to
converge.  Therefore, neither by analytical nor by numerical means will one
be able to compute a useful semiclassical signal from (\ref{C1}).  The
inclusion of uniform approximations in semiclassical quantization is thus a
nontrivial and challenging task.

To solve the problem we resort to the observation made in \cite{Mai00} that
a band-limited signal, which only contains the spectral information
describing the quantum system in a finite energy interval $[E_{\rm min},
E_{\rm max}]$, can be obtained by restricting the energy integral in
(\ref{C1}) to this window.  The resulting signal
\begin{equation}
\label{CBL}
  C^{\rm bl}(t) = \frac{1}{2\pi} \int_{E_{\rm min}}^{E_{\rm max}} 
    g^{\rm osc}(E) e^{-iEt} dE \; ,
\end{equation}
where $g^{\rm osc}(E)$ is in general a non-scaling function that includes
both contributions of isolated closed orbits and uniform approximations,
can be evaluated numerically by standard methods. As it contains only a
small number of frequencies in the interval chosen, it can be processed in
a numerically stable way by conventional high-resolution methods such as
linear prediction or Pad\'e approximants.  Contrary to previous methods for
scaling systems, all of which contained the analytic evaluation of an
integral, the numerical integration imposes no restrictions on the
semiclassical response function occurring in the integrand. Notice that our
method cannot be applied in connection with the original filter
diagonalization algorithm \cite{Mai99d}. It is only the separation between
a low-resolution frequency filtering stage and a high-resolution harmonic
inversion stage introduced in \cite{Mai00} that allows for the present
generalization to arbitrary non-scaling semiclassical signals.

To demonstrate our method we investigate Stark spectra of the hydrogen atom
for transitions from the ground state $|1s0\rangle$ to highly excited
Rydberg states with light polarized parallel to the electric field axis.
The external field strength is
$F=10^{-8}\,\text{a.u.}\ \hat{=}\ 51.4\,\text{V/cm}$.  The high-resolution
Stark spectrum is obtained in two steps:

\begin{figure}
  \includegraphics[width=0.9\columnwidth]{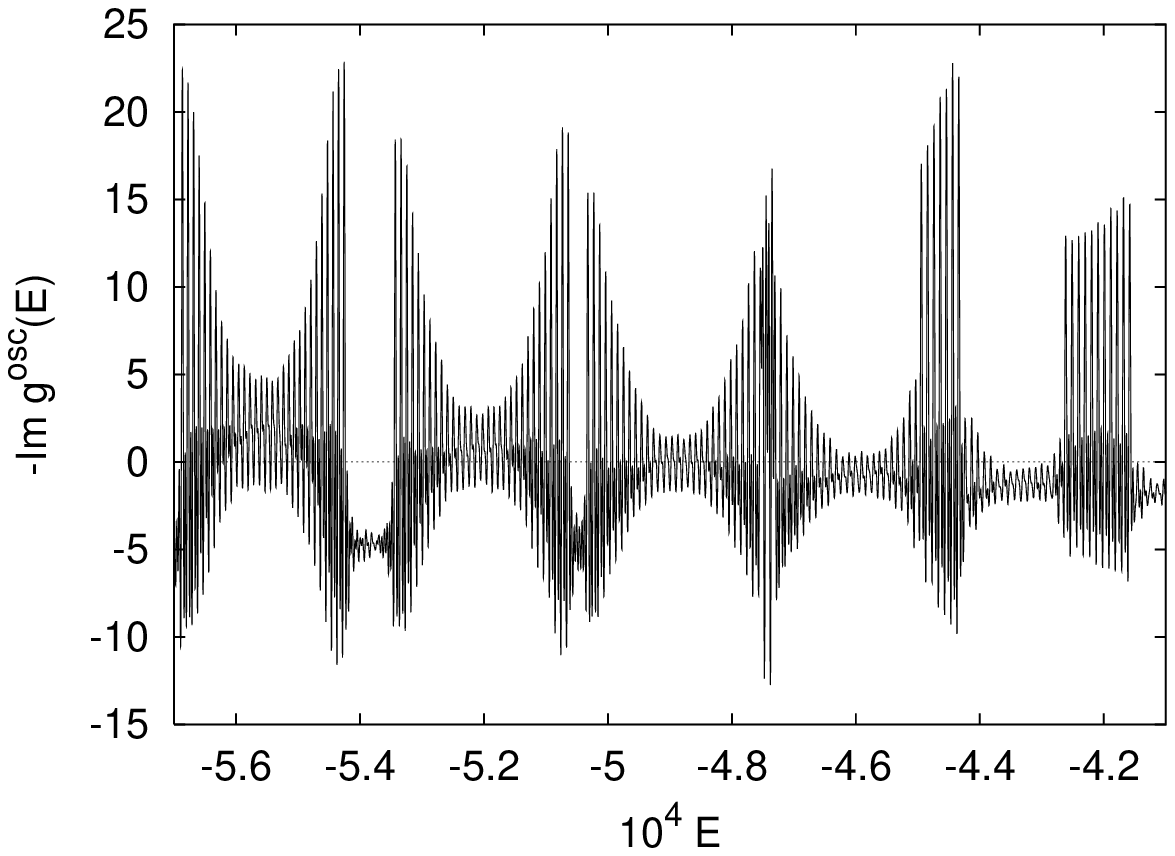}
  \includegraphics[width=0.9\columnwidth]{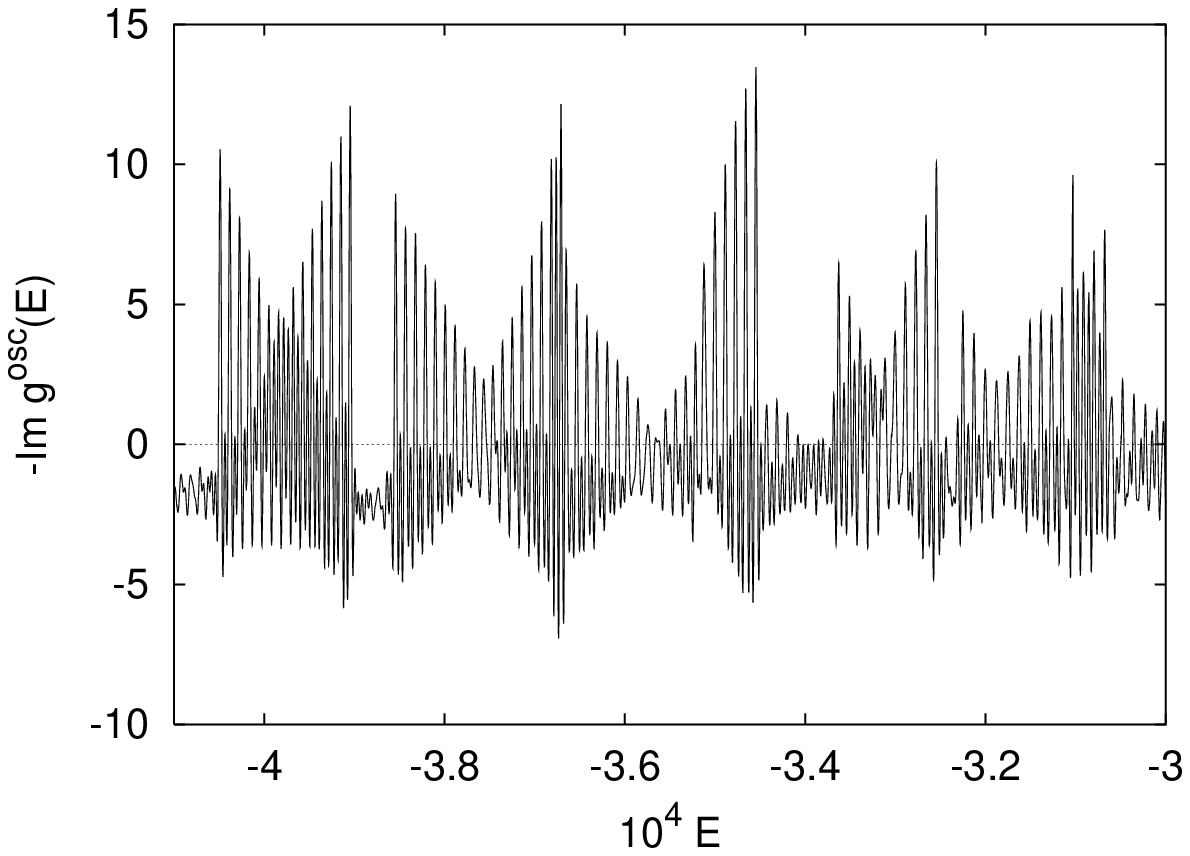}
  \caption{Low-resolution semiclassical photo-absorption spectrum for the
  hydrogen atom in an electric field $F=51.4$~V/cm with initial state 
  $|1s0\rangle$ and light polarized along the electric field axis.
  The scaled truncation time is $T_{\rm max}=15\times 10^6$.}
\label{fig:lores}
\end{figure}

\begin{figure}
  \includegraphics[width=0.9\columnwidth]{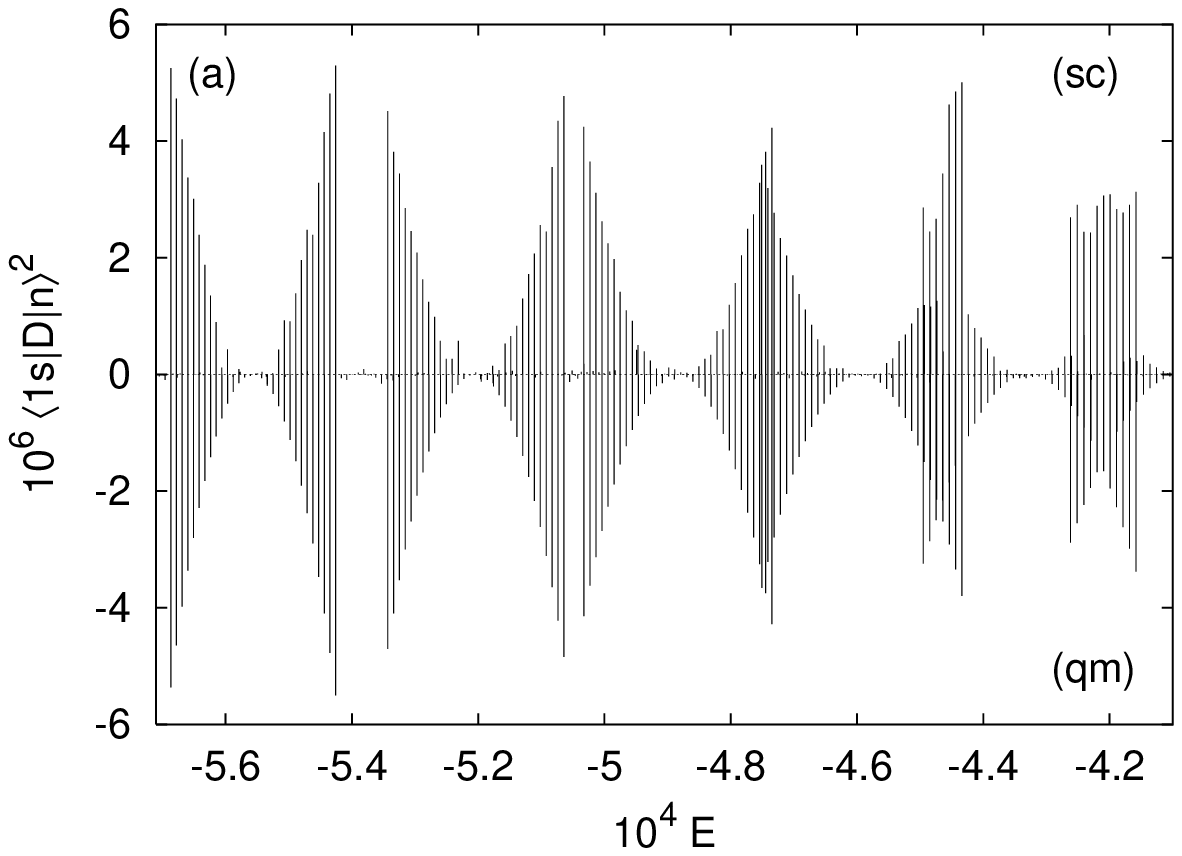}
  \includegraphics[width=0.9\columnwidth]{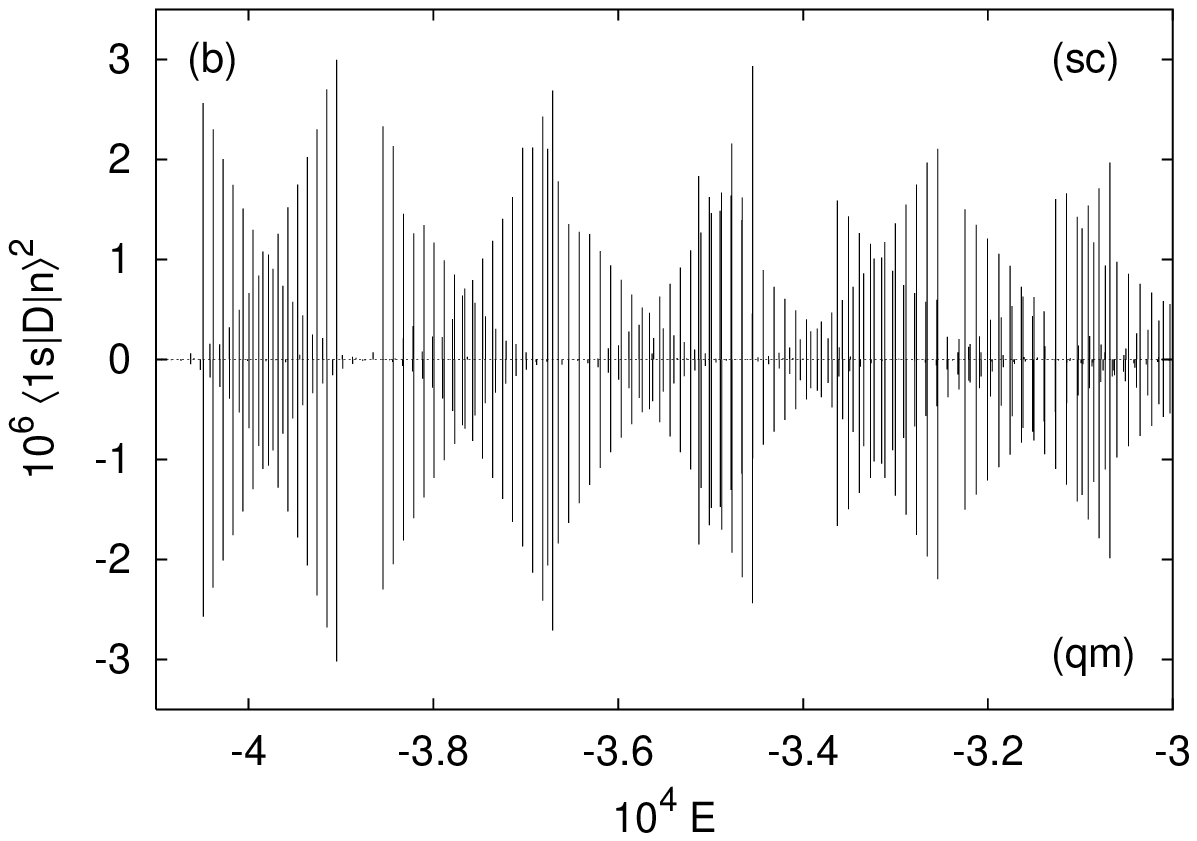}
  \caption{High-resolution semiclassical (upper part) and quantum (lower
  part, inverted) photo-absorption spectrum for the hydrogen atom in an
  electric field $F=51.4$~V/cm with initial state $|1s0\rangle$ and light
  polarized along the electric field axis. The truncation time for the
  semiclassical spectrum is $T_{\rm max}=40\times 10^6$.}
\label{fig:hires}
\end{figure}

Firstly, a low-resolution semiclassical spectrum is obtained by truncating 
the closed-orbit sum (\ref{COSum}) at a maximal period $T_{\rm max}$.
The cut-off value for the periods should not be chosen smaller than
the signal length of the band-limited signal (\ref{CBL}) used for the
harmonic inversion in the second step of the procedure.
The low-resolution spectrum calculated with a scaled truncation time 
of $T_{\rm max}=15\times 10^6$ is presented in Fig.~\ref{fig:lores}.
In the lower energy range shown in Fig.~\ref{fig:lores}(a), individual 
non-overlapping $n$-manifolds can be observed.
(We have $n=30$ at $E\approx-5.56\times 10^{-4}$.) 
In this region, the signal is sufficiently long to resolve individual spectral
lines, although their precise determination from the plots remains difficult.
In the higher energy range shown in Fig.~\ref{fig:lores}(b), two, three, or 
even four different $n$-manifolds overlap, leading to a drastically increased
spectral density.  In this region the semiclassical signal is evidently too
short to discriminate individual lines. 
It is important to note that the possibility of computing the low-resolution 
spectrum depends critically on the use of uniform approximations. 
If it was calculated from isolated-orbit contributions only, a dense 
sequence of bifurcation-induced divergences would cover even the 
large-scale structure of the spectra.
%
% calculation of the band-limited time signal, T_max=40 !
With the low-resolution semiclassical spectrum (Fig.~\ref{fig:lores}) at hand
the band-limited time signal (\ref{CBL}) is now obtained by a numerical
Fourier transform of $g^{\rm osc}(E)$.
In that calculation we used the signal length $T_{\rm max}=40\times 10^6$ 
in order to resolve individual levels in the region of overlapping 
$n$-manifolds.

In the second step, the high-resolution semiclassical spectrum is finally
obtained by harmonic inversion of the band-limited time signal (\ref{CBL}).
In Fig.~\ref{fig:hires} the semiclassical spectrum is compared to the 
exact quantum spectrum.
The overall agreement between the semiclassical and the quantum spectrum is
excellent, although for a few levels the comparison reveals discrepancies
between the semiclassical and the quantum matrix elements.  Note, in
particular, the region of high spectral density at $E\approx
3.2\times10^{-4}$. In this region, groups of 3 nearly degenerate levels
exist, some of which are well resolved semiclassically.  At $E\approx
-4.4\times 10^{-4}$, even closer lines exist -- they can hardly be
discerned in the quantum spectrum.  
These lines are not resolved semiclassically.
Instead, the harmonic inversion yields single lines with amplitudes equal to 
the sum of the two quantum amplitudes.
We are confident to fully resolve even these states in the semiclassical 
spectrum when applying the cross-correlation technique for harmonic 
inversion \cite{Mai99d}.

In the future it should also be possible to extend the semiclassical spectrum
to energies $E>E_{\rm S}$ where the classical motion is not completely 
bound, and to extract the semiclassical widths of the Stark resonances.
A particular challenge is posed by the region around the Stark saddle point
energy $E_{\rm S}$. Before the downhill orbit ceases to exist at $E_{\rm
S}$, it undergoes an infinite sequence of bifurcations, giving birth to
non-axial orbits with arbitrarily high uphill repetition numbers. If
subsequent bifurcations of a single orbit are too close, the uniform
approximation (\ref{uniform:eq}) is no longer appropriate. It must then be
replaced with a uniform approximation describing several bifurcations
collectively. The uniformization of an infinite bifurcation cascade, in
particular, remains an open problem whose solution is required to
semiclassically cross the saddle point energy.

% Because the exact number of frequencies contained in a signal is not known
% \emph{a priori}, the harmonic inversion technique yields some spurious
% spectral lines. It also gives an error parameter to distinguish between
% true and spurious eigenvalues. For the spectra presented here, no selection
% of ``good'' frequencies has been done except with respect to the error
% parameter, i.e., spectral lines have not been selected according to their
% agreement with the quantum results. A few spurious lines can therefore be
% seen in the semiclassical spectrum, all of which, however, have got very
% low amplitudes. This observation confirms that the semiclassical
% approximation reproduces the quantum response function well.
% 
% Even with the new method at hand, the quantization of a non-scaling system
% is still harder than the calculation of a scaled spectrum because the
% classical orbits must be calculated across an energy rather than at a
% single scaled energy. This disadvantage, however, is inherent in the
% endeavor to semiclassically approximate the quantum response
% function. It is therefore not due to limitations of the harmonic inversion
% procedure, but a difficulty any quantization scheme has to face.

In summary, we have extended the harmonic inversion approach to
semiclassical quantization to the quantization of systems without a scaling
property.
The generalized method allows for the inclusion of uniform approximations 
into the quantization procedure.  We have
demonstrated the effectiveness of our method by calculating a high-quality
semiclassical
spectrum for the hydrogen atom in an electric field in a spectral region
where the semiclassical approximation without uniform approximations would
be completely useless due to the abundance of bifurcations. With the
modifications presented here, the technique of quantization by harmonic
inversion has reached a stage where it does not impose any conditions on
the classical dynamics of the system under study except that a semiclassical 
approximation to the response function can be given. Besides
uniform approximations, any other non-standard semiclassical contributions 
such as diffractive corrections can be included.
Thus, the harmonic inversion can now be regarded as a truly universal tool 
for the semiclassical quantization of arbitrary systems.

%\bibliography{paper}

\begin{thebibliography}{16}
\expandafter\ifx\csname natexlab\endcsname\relax\def\natexlab#1{#1}\fi
\expandafter\ifx\csname bibnamefont\endcsname\relax
  \def\bibnamefont#1{#1}\fi
\expandafter\ifx\csname bibfnamefont\endcsname\relax
  \def\bibfnamefont#1{#1}\fi
\expandafter\ifx\csname citenamefont\endcsname\relax
  \def\citenamefont#1{#1}\fi
\expandafter\ifx\csname url\endcsname\relax
  \def\url#1{\texttt{#1}}\fi
\expandafter\ifx\csname urlprefix\endcsname\relax\def\urlprefix{URL }\fi
\providecommand{\bibinfo}[2]{#2}
\providecommand{\eprint}[2][]{\url{#2}}

\bibitem[{\citenamefont{Gutzwiller}(1990)}]{Gut90}
\bibinfo{author}{\bibfnamefont{M.~C.} \bibnamefont{Gutzwiller}},
  \emph{\bibinfo{title}{Chaos in Classical and Quantum Mechanics}}
  (\bibinfo{publisher}{Springer}, \bibinfo{address}{New York},
  \bibinfo{year}{1990}).

\bibitem[{\citenamefont{Du and Delos}(1988)}]{Du88}
\bibinfo{author}{\bibfnamefont{M.~L.} \bibnamefont{Du}} \bibnamefont{and}
  \bibinfo{author}{\bibfnamefont{J.~B.} \bibnamefont{Delos}},
  \bibinfo{journal}{Phys.~Rev.~A} \textbf{\bibinfo{volume}{38}},
  \bibinfo{pages}{1896 and 1913} (\bibinfo{year}{1988}).

\bibitem[{\citenamefont{Bogomolny}(1989)}]{Bog89}
\bibinfo{author}{\bibfnamefont{E.~B.} \bibnamefont{Bogomolny}},
  \bibinfo{journal}{Sov.~Phys.~JETP} \textbf{\bibinfo{volume}{69}},
  \bibinfo{pages}{275} (\bibinfo{year}{1989}).

\bibitem[{\citenamefont{Main et~al.}(1994)\citenamefont{Main, Wiebusch, Welge,
  Shaw, and Delos}}]{Mai94}
\bibinfo{author}{\bibfnamefont{J.}~\bibnamefont{Main}},
  \bibinfo{author}{\bibfnamefont{G.}~\bibnamefont{Wiebusch}},
  \bibinfo{author}{\bibfnamefont{K.}~\bibnamefont{Welge}},
  \bibinfo{author}{\bibfnamefont{J.}~\bibnamefont{Shaw}}, \bibnamefont{and}
  \bibinfo{author}{\bibfnamefont{J.~B.} \bibnamefont{Delos}},
  \bibinfo{journal}{Phys.~Rev.~A} \textbf{\bibinfo{volume}{49}},
  \bibinfo{pages}{847} (\bibinfo{year}{1994}).

\bibitem[{\citenamefont{Courtney et~al.}(1995)\citenamefont{Courtney, Jiao,
  Spellmeyer, Kleppner, Gao, and Delos}}]{Cou95}
\bibinfo{author}{\bibfnamefont{M.}~\bibnamefont{Courtney}},
  \bibinfo{author}{\bibfnamefont{H.}~\bibnamefont{Jiao}},
  \bibinfo{author}{\bibfnamefont{N.}~\bibnamefont{Spellmeyer}},
  \bibinfo{author}{\bibfnamefont{D.}~\bibnamefont{Kleppner}},
  \bibinfo{author}{\bibfnamefont{J.}~\bibnamefont{Gao}}, \bibnamefont{and}
  \bibinfo{author}{\bibfnamefont{J.~B.} \bibnamefont{Delos}},
  \bibinfo{journal}{Phys.~Rev.~Lett.} \textbf{\bibinfo{volume}{74}},
  \bibinfo{pages}{1538} (\bibinfo{year}{1995}).

\bibitem[{\citenamefont{Kips et~al.}(1999)\citenamefont{Kips, Vassen, and
  Hogervorst}}]{Kip99}
\bibinfo{author}{\bibfnamefont{A.}~\bibnamefont{Kips}},
  \bibinfo{author}{\bibfnamefont{W.}~\bibnamefont{Vassen}}, \bibnamefont{and}
  \bibinfo{author}{\bibfnamefont{D.~W.} \bibnamefont{Hogervorst}},
  \bibinfo{journal}{Phys.~Rev.~A} \textbf{\bibinfo{volume}{59}},
  \bibinfo{pages}{2948} (\bibinfo{year}{1999}).

\bibitem[{\citenamefont{Main et~al.}(1997)\citenamefont{Main, Mandelshtam, and
  Taylor}}]{Mai97c}
\bibinfo{author}{\bibfnamefont{J.}~\bibnamefont{Main}},
  \bibinfo{author}{\bibfnamefont{V.~A.} \bibnamefont{Mandelshtam}},
  \bibnamefont{and} \bibinfo{author}{\bibfnamefont{H.~S.}
  \bibnamefont{Taylor}}, \bibinfo{journal}{Phys.~Rev.~Lett.}
  \textbf{\bibinfo{volume}{79}}, \bibinfo{pages}{825} (\bibinfo{year}{1997}).

\bibitem[{\citenamefont{Main}(1999)}]{Mai99d}
\bibinfo{author}{\bibfnamefont{J.}~\bibnamefont{Main}},
  \bibinfo{journal}{Physics Reports} \textbf{\bibinfo{volume}{316}},
  \bibinfo{pages}{233} (\bibinfo{year}{1999}).

\bibitem[{\citenamefont{Main and Wunner}(1997)}]{Mai97a}
\bibinfo{author}{\bibfnamefont{J.}~\bibnamefont{Main}} \bibnamefont{and}
  \bibinfo{author}{\bibfnamefont{G.}~\bibnamefont{Wunner}},
  \bibinfo{journal}{Phys.~Rev.~A} \textbf{\bibinfo{volume}{55}},
  \bibinfo{pages}{1743} (\bibinfo{year}{1997}).

\bibitem[{\citenamefont{Gao and Delos}(1997)}]{Gao97}
\bibinfo{author}{\bibfnamefont{J.}~\bibnamefont{Gao}} \bibnamefont{and}
  \bibinfo{author}{\bibfnamefont{J.~B.} \bibnamefont{Delos}},
  \bibinfo{journal}{Phys.~Rev.~A} \textbf{\bibinfo{volume}{56}},
  \bibinfo{pages}{356} (\bibinfo{year}{1997}).

\bibitem[{\citenamefont{Gao and Delos}(1994)}]{Gao94}
\bibinfo{author}{\bibfnamefont{J.}~\bibnamefont{Gao}} \bibnamefont{and}
  \bibinfo{author}{\bibfnamefont{J.~B.} \bibnamefont{Delos}},
  \bibinfo{journal}{Phys.~Rev.~A} \textbf{\bibinfo{volume}{49}},
  \bibinfo{pages}{869} (\bibinfo{year}{1994}).

\bibitem[{\citenamefont{Kondratovich and Delos}(1997)}]{Kon97}
\bibinfo{author}{\bibfnamefont{V.}~\bibnamefont{Kondratovich}}
  \bibnamefont{and} \bibinfo{author}{\bibfnamefont{J.~B.} \bibnamefont{Delos}},
  \bibinfo{journal}{Phys.~Rev.~A} \textbf{\bibinfo{volume}{56}},
  \bibinfo{pages}{R5} (\bibinfo{year}{1997}); 
  \textbf{\bibinfo{volume}{57}},
  \bibinfo{pages}{4654} (\bibinfo{year}{1998}).

\bibitem[{\citenamefont{Gao et~al.}(1992)\citenamefont{Gao, Delos, and
  Baruch}}]{Gao92}
\bibinfo{author}{\bibfnamefont{J.}~\bibnamefont{Gao}},
  \bibinfo{author}{\bibfnamefont{J.~B.} \bibnamefont{Delos}}, \bibnamefont{and}
  \bibinfo{author}{\bibfnamefont{M.}~\bibnamefont{Baruch}},
  \bibinfo{journal}{Phys.~Rev.~A} \textbf{\bibinfo{volume}{46}},
  \bibinfo{pages}{1449} (\bibinfo{year}{1992}).

\bibitem[{\citenamefont{Shaw and Robicheaux}(1998)}]{Shaw98a}
\bibinfo{author}{\bibfnamefont{J.~A.} \bibnamefont{Shaw}} \bibnamefont{and}
  \bibinfo{author}{\bibfnamefont{F.}~\bibnamefont{Robicheaux}},
  \bibinfo{journal}{Phys.~Rev.~A} \textbf{\bibinfo{volume}{58}},
  \bibinfo{pages}{1910} (\bibinfo{year}{1998}).

\bibitem[{\citenamefont{Abramowitz and Stegun}(1984)}]{Abramowitz}
\bibinfo{author}{\bibfnamefont{M.}~\bibnamefont{Abramowitz}} \bibnamefont{and}
  \bibinfo{author}{\bibfnamefont{I.~A.} \bibnamefont{Stegun}},
  \emph{\bibinfo{title}{Pocketbook of Mathematical Functions}}
  (\bibinfo{publisher}{Verlag Harri Deutsch},
  \bibinfo{address}{Frankfurt/Main}, \bibinfo{year}{1984}).

\bibitem[{\citenamefont{Main et~al.}(2000)\citenamefont{Main, Dando,
  Belki{\'c}, and Taylor}}]{Mai00}
\bibinfo{author}{\bibfnamefont{J.}~\bibnamefont{Main}},
  \bibinfo{author}{\bibfnamefont{P.~A.} \bibnamefont{Dando}},
  \bibinfo{author}{\bibfnamefont{D\v z.}~\bibnamefont{Belki{\'c}}},
  \bibnamefont{and} \bibinfo{author}{\bibfnamefont{H.~S.}
  \bibnamefont{Taylor}}, \bibinfo{journal}{J.~Phys.~A}
  \textbf{\bibinfo{volume}{33}}, \bibinfo{pages}{1247} (\bibinfo{year}{2000}).

\end{thebibliography}

\end{document}